\documentclass[a4paper,12pt]{article}
\usepackage[utf8x]{inputenc}
\usepackage{amsmath,amssymb,cite}
\usepackage{hyperref}

\numberwithin{equation}{section}

\title{{\bf Entropy of Quantum States: Ambiguities}}
\author{A. P. Balachandran$^{a,b}$\footnote{balachandran38@gmail.com},\,, A. R. de Queiroz$^{c}$\footnote{amilcarq@unb.br}\, and S. Vaidya$^d$\footnote{vaidya@cts.iisc.ernet.in} \\
\begin{small}{\it $^a$Department of Physics, Syracuse University, Syracuse, N. Y. 13244-1130, USA}
\end{small}\\
\begin{small}{\it $^b$Institute of Mathematical Sciences, Chennai, India} \end{small} \\
\begin{small}{\it $^c$Instituto de Fisica, Universidade de Brasilia, Caixa Postal 04455, 70919-970, Brasilia, DF, Brazil} \end{small}\\
\begin{small}{\it $^d$Centre for High Energy Physics, Indian Institute of Science, Bangalore, 560012, India}
\end{small}}

\date{\empty}

\begin{document}

\maketitle

\begin{abstract}
The von Neumann entropy of a generic quantum state is not unique unless the state can be uniquely 
decomposed as a sum of extremal or pure states. As pointed out to us by Sorkin, this happens if the GNS 
representation (of the algebra of observables in some quantum state) is reducible, and some representations 
in the decomposition occur with non-trivial degeneracy. This non-unique entropy can occur at zero 
temperature. We will argue elsewhere in detail that the degeneracies in the GNS 
representation can be interpreted as an emergent broken gauge symmetry, and play an important role in 
the analysis of emergent entropy due to non-Abelian anomalies.
Finally, we establish the analogue of an H-theorem for this entropy by showing that its evolution is 
Markovian, determined by a stochastic matrix. 
\end{abstract}

\section{Introduction}

For a quantum system with density matrix $\rho$, a common definition of the von Neumann entropy is 
\begin{equation}
S(\rho) = - {\rm Tr}\, \rho \log \rho.
\label{vNentropy}
\end{equation}
There is also another definition for the entropy which goes as follows. Let $\rho_i$ be "pure" or 
rank 1 density matrices. Since any density matrix is a convex combination of $\rho_i$, we can write
\begin{equation}
\rho = \sum_i \lambda_i \rho_i, \quad \lambda_i \geq 0, \quad \sum_i \lambda_i = 1.
\label{densitymatrix}
\end{equation}
Then 
\begin{equation}
S(\rho) = -\sum_i \lambda_i \log \lambda_i
\label{entropydef}
\end{equation}
The interpretation of $\lambda_i$ is that it is the probability of finding the pure state described by $\rho_i$ in the 
state described by $\rho$. This definition is close to the definition used in classical statistical mechanics.

Now, two density matrices $\rho'$ and $\rho''$ giving the same expectation value for observables define the 
{\it same state $\omega$}: as a state $\omega$, they must be identified. [See later for the definition of a state 
$\omega$ without using density matrices.] But in case $\rho' \neq \rho''$, the two definitions above can give differing entropies, 
$S(\rho') \neq S(\rho'')$. 

A sufficient condition to associate a unique entropy to a state is that the representation of a state by a 
density matrix is unique: $\rho' = \rho''$. In other words, a sufficient condition to associate a unique entropy 
to a state is that it is {\it uniquely} representable in terms of rank 1 density matrices, i.e. pure or extremal states. 

Now recall that the set of states forms a convex set: if $\omega', \omega''$ are states, then 
$\lambda \omega' + (1-\lambda) \omega'', \lambda \in [0,1]$ is also a state. For the unique association of 
entropy to a state, the above condition means that this convex set is also a ``simplex". Any point of a simplex 
is uniquely representable in terms of its extremal points. Standard simplices like a triangle or a tetrahedron 
have this property. 

But the set of quantum states is rarely a simplex. Already for a qubit with ${\mathbb C}^2$ as Hilbert space,
the set of states is the three-dimensional ball ${\mathcal B}^3$ with pure states being its boundary 
$\partial {\mathcal B}^3 = S^2 \simeq {\mathbb C}P^1$. The number of pure states is infinite, being 
${\mathbb C}P^1$ as a (complex) manifold. It is easy to see that a generic state in the interior of ${\mathcal B}^3$ 
allows many representations like (\ref{densitymatrix}) with differing coefficients $\lambda_i$ and different 
values of $S(\rho)$.

This point was emphasized by Sorkin \cite{sorkin_pc} to one of the authors (APB). It is also known in the 
literature \cite{Schroer:1998ey}.

But not all quantum systems fail to have simplices as quantum states. An important exemption is the set of 
KMS states of a given family (with the same `modular Hamiltonian'). These do form a simplex, a result 
with important implications for phase transitions \cite{haagbook}. Another class, not necessarily disjoint 
from above, occurs already in finite dimensions. It arises as follows. 

Given a state $\omega$ on a ($C^*$-) algebra ${\cal A}$ of observables, there is a canonical construction of 
a Hilbert space ${\cal H}_\omega$ which carries a representation $\pi_\omega$ of ${\cal A}$. It is called the 
GNS construction and need not be irreducible. In any case ${\cal H}_\omega$ admits an orthogonal direct 
sum decomposition 
\begin{equation}
{\cal H}_\omega =\bigoplus_{r,j} {\cal H}^{(r,j)}_\omega
\label{hilbertdecomp}
\end{equation}
where ${\cal H}^{(r,j)}_\omega$ carries the irreducible representation $\pi^{(r)}_\omega$ and $j$ is a degeneracy 
index for the representation. If there is no degeneracy so that $j$ assumes only one value for each $r$, then 
the state $\omega$ has a unique representation in terms of extremal states or rank 1 density matrices. 
The convex set of states generated by these extremals do form a simplex. As a corollary, we have a 
unique value for entropy. 

But if there is degeneracy for any $r$, then, as we shall see, the convex set generated by the extremals is not 
a simplex. There are many ways of writing $\omega$ in terms of pure states and there is no unique entropy for 
$\omega$. 

We will accept (\ref{vNentropy}) or  (\ref{entropydef}) as the correct definition of the entropy. For us, they 
are equivalent since we will work with $\rho_i$ such that ${\rm Tr} \rho_i \rho_j = \delta_{ij}$. They have 
the correct physical interpretation. It appears that the uncertainties in the value of entropy generically arise 
from degeneracies in the GNS representation $\pi_\rho$ as indicated above and the consequent failure of 
the set of states to be a simplex. Sorkin has emphasized this point in his recent work on black hole entropy 
\cite{Sorkin:2012sn,Afshordi:2012jf,Afshordi:2012ez} and pointed it out to us as well. 

In this paper, after reviewing earlier work on the GNS construction, we will illustrate the emergence 
of degeneracies already when ${\cal A}$ is the $2 \times 2$ matrix algebra. An important contribution here is 
the interpretation of the degeneracy in terms of an emergent broken ``gauge symmetry''. 

We also establish a sort of $H$-theorem for $S(\rho)$. Thus the change of $\lambda_i$'s to another set 
$\lambda'_i$'s is a Markov process, determined by a stochastic matrix. Then since $S(\rho)$ is a concave function 
$\hat{S}(\lambda)$ of $\lambda = (\lambda_1, \lambda_2, \cdots,)$, one proves that 
\begin{equation}
\hat{S}(\lambda') \geq \hat{S}(\lambda)
\end{equation}

The observation that $\hat{S}$ is a non-decreasing function of $\lambda$ for a Markovian evolution governed 
by a stochastic matrix is also due to Sorkin \cite{Sorkin:1997ja}. Our observation that different extremal decompositions of $\rho$ lead to such 
a Markovian evolution of $\lambda$ is new. We find that 
\begin{equation}
\hat{S} (\lambda'_0) = \hat{S}(\lambda_0)
\end{equation}
if and only if $\lambda_0$ is a maximum of $S$, which occurs when the system is maximally disordered. Then
\begin{equation}
\lambda'_{0i} = \lambda_{0i}
\end{equation}
is independent of $i$.

We note that such entropy is in principle measurable. What is quoted above is the entropy per particle. That 
will get enhanced in a gas of these particles and affect, say, the specific heat.

Another point to note is that what we quote is the zero temperature entropy. So it does not vanish or necessarily have 
a unique value as happens in quantum phase transitions. Our observations on the non-uniqueness of entropy may have striking consequences for studies of quantum phase transitions. 

Now, in the partial tracing approach, there is a result which shows that any mixed state can be 'purified'  or 
made extremal by tensoring the given Hilbert space with another, the given mixed state then being obtaining 
by partial tracing. There is a similar result in our approach as well. That involves enhancing 
the physical subalgebra ${\cal A}$ by the ``gauge algebra'' ${\cal A'}$ (which commutes with ${\cal A}$) and 
working with ${\cal A} \otimes {\cal A'}$. We will also see how this works towards the end. This construction
will play a crucial role in the analysis of emergent entropy due to non-abelian anomalies. Elsewhere \cite{bqv}, 
we will discuss this in the context of ethylene molecule (where the double cover $D^*_8$ of the dihedral group  
plays the role of the non-abelian ``gauge symmetry'' group) and color symmetry breaking in gauge theories 
due to non-abelian monopoles.

\section{The GNS Construction}

The GNS construction shows that the Hilbert space formulation of quantum theory is an emergent 
construction from a state $\omega$ on an algebra ${\cal A}$. Both the state and the algebra can be 
presented abstractly in this fundamental approach. 

The algebra ${\cal A}$ represents the algebra of observables. There is a $*$-operation 
$\alpha \rightarrow \alpha^*$ on its elements $\alpha \in {\cal A}$ which is an anti-linear involution: 
$\alpha^{**} = \alpha$. (In a ``$*$''-representation $\pi$ of ${\cal A}$, $*$ becomes $\dagger$, the 
hermitean conjugation: $\pi(\alpha^*) = \pi(\alpha)^\dagger$.) ${\cal A}$ contains unity ${\bf 1}$. Finally there is 
a norm $\| \cdot \|$ defined on elements of ${\cal A}$ fulfilling $\|\alpha^* \alpha \| = \|\alpha\|^2$. This makes 
${\cal A}$ into a $C^*$-algebra (after it is completed in the norm $\|\cdot\|$). 

All $N\times N$ matrix algebras $M_N$ are $C^*$ where $*$ is the hermitean conjugation and $\|m\|^2$ for 
$m \in M_N$ is the largest eigenvalue of $m^\dagger m$. 

The concept of a state $\omega$ on ${\cal A}$ is abstracted from what we know of density matrices $\rho$. 
Although often the distinction between $\rho$ and $\omega$ is not maintained in the literature, they are 
really different: whereas $\rho$ is a linear operator, $\omega$ is not. So henceforth we will distinguish a 
state from a density matrix.

The definition of a state $\omega$ on ${\cal A}$ is as follows. It is a linear map from ${\cal A}$ to 
complex numbers fulfilling 
\begin{equation}
\omega({\bf 1}) = 1, \quad \omega(\alpha^*) = \overline{\omega(\alpha)}, \quad \omega(\alpha^* \alpha) 
\geq 0,
\label{statedef}
\end{equation}
the bar denoting complex conjugation. 

A density matrix $\rho$ naturally defines a state $\omega_\rho$ as follows. On the Hilbert space ${\cal H}$ 
of the operator $\rho$, there is a representation $\pi$ of ${\cal A}$ which gives observables as operators. 
Then
\begin{equation}
\omega_\rho(\alpha) = {\rm Tr}\, \rho \pi (\alpha).
\end{equation}

Thus the operator $\rho$ defines a state $\omega_\rho$. Conversely, we will see below that a state 
defines at least one density matrix fulfilling (\ref{statedef}).

But in general, many density matrices may give the same state. So the map from states to density matrices need not be unique. It will not be under conditions leading to 
(\ref{hilbertdecomp}), with $j$ taking more than one value. We will see this below.

Here is the GNS construction. As in the discussion of regular representations of groups, we associate a 
vector space $\tilde{{\cal A}}$ with elements $|\alpha \rangle$ for each $\alpha$ belonging to ${\cal A}$. It 
has obvious linearity properties: if $\alpha, \beta \in {\cal A}$ and $ \lambda, \mu \in {\mathbb C}$, then 
$|\lambda \alpha + \mu \beta \rangle = \lambda |\alpha \rangle + \mu |\beta \rangle$.

The next step is to put an inner product on these kets using the given state $\omega$ on ${\cal A}$. We set
\begin{equation}
\langle \alpha | \beta \rangle = \omega (\beta^* \alpha).
\end{equation}
This is an inner product,
\begin{equation}
\langle \alpha|\alpha \rangle = \omega(\alpha^* \alpha) \geq 0
\end{equation}
but not a scalar product since $\langle \alpha | \beta \rangle$ may be 0 for non-zero $|\alpha \rangle$. 
Let ${\cal N}_\omega$ denote the subspace of ${\cal A}$ which maps to the space 
$\widetilde{{\cal N}_\omega}$ of these null vectors. 

Now like all inner products, the above also fulfills Schwarz inequality:
\begin{equation}
|\langle \beta|\alpha \rangle|^2 \leq \langle \beta| \beta \rangle \langle \alpha | \alpha \rangle.
\end{equation}
From this follows that $\widetilde{{\cal N}_\omega}$ is orthogonal to all vectors of $\tilde{{\cal A}}$:
\begin{equation}
|\langle \alpha | n \rangle|^2 \leq \langle \alpha | \alpha \rangle \langle n | n \rangle = 0, 
\quad \forall \alpha \in {\cal A}, \,\, n \in {\cal N}_\omega.
\end{equation}

The next step is to introduce a representation ${\cal A}_L, \alpha \rightarrow \alpha_L$ on $\tilde{{\cal A}}$.
It is the analog of the left-regular representation of groups and is defined by
\begin{equation}
\alpha_L | \beta \rangle = |\alpha \beta \rangle.
\end{equation}
It follows that ${\cal N}_\omega$ is a left ideal:
\begin{equation}
\langle \beta | \alpha n \rangle = \langle \alpha^* \beta | n \rangle =0 \quad 
{\rm if} \quad \langle n | n \rangle =0.
\end{equation}
Thus all vectors $\widetilde{{\cal N}_\omega} = \{ |n \rangle \}, n \in {\cal N}_\omega$ span an invariant 
subspace under ${\cal A}_L = \{\alpha_L \}$. They are all null and orthogonal to all vectors of $\tilde{\cal A}$.
So consider 
\begin{equation}
{\cal H}^0_\omega = \tilde{\cal A}/\widetilde{\cal N_\omega}= \{|[\alpha] \rangle, [\alpha]:= \alpha + 
{\cal N_\omega} \}
\end{equation}
with the scalar product 
\begin{equation}
\langle [\beta] | [\alpha] \rangle = \omega(\beta^* \alpha) 
\end{equation}
for any two choices $\beta,\alpha$ from the equivalence classes $[\beta],[\alpha]$. The right-hand side 
is independent of this choice. $|{\cal N}_\omega \rangle$ acts as the zero vector of ${\cal H}^0_\omega$.
There are no other null vectors.

Finally, ${\cal H}^0_\omega$ carries a representation $\pi_\omega$ of ${\cal A}$:
\begin{equation}
\pi_\omega (\alpha) |[\beta] \rangle = |[\alpha \beta] \rangle
\end{equation}
Here we use the fact that $\alpha {\cal N_\omega} = {\cal N_\omega}$. 

We can now complete ${\cal H}^0_\omega$ into a Hilbert space ${\cal H}_\omega$. 

Finally we have a representation $\pi_\omega$ of ${\cal A}$ on ${\cal H}_\omega$. This is the 
GNS representation.

Note that if
\begin{equation}
\rho_\omega = |[{\bf 1}] \rangle \langle [{\bf 1}]|,
\end{equation}
then
\begin{equation}
{\rm Tr}\, [\rho_\omega \pi_\omega(\alpha)] = \langle [{\bf 1}]|[\alpha] \rangle = \omega(\alpha)
\end{equation}
so that $\rho_\omega$ is a density matrix associated with $\omega$.

The representation $\pi_\omega$ on ${\cal H}_\omega$ need not be irreducible. But we can always write it as a 
direct sum of irreducible components as in (\ref{hilbertdecomp}).

\section{An Example: $M_2(\mathbb{C})$}

This example is adapted from \cite{Balachandran:2012rw} and illustrates the GNS construction. 

A basis from $M_2(\mathbb{C})$ are the ``matrix units'' $e_{ij}, \,(i,j = 1,2)$ where
\begin{equation}
(e_{ij})_{kl} = \delta_{ik}\delta_{jl}.
\end{equation}
They fulfill
\begin{equation}
e_{ij} e_{kl} = \delta_{jk} e_{il}.
\end{equation}
It is convenient to write $e_{ij}$ in Dirac's notation:
\begin{equation}
e_{ij} = |i \rangle \langle j |.
\end{equation}
A general element $\alpha \in M_2(\mathbb{C})$ can be written as 
\begin{equation}
\alpha = \sum_{i,j} \alpha_{ij} |i \rangle \langle j |, \quad \alpha_{ij} \in \mathbb{C}.
\end{equation}

Now for the state. We consider a family of $\omega_\lambda \,(\lambda \in [0,1])$ of these states and 
define them abstractly and not as density matrices:
\begin{equation}
\omega_\lambda (\alpha) = \lambda \alpha_{11} + (1-\lambda) \alpha_{22}.
\end{equation}
Clearly,
\begin{equation}
\omega_\lambda ({\bf 1}) = 1,
\end{equation}
and since
\begin{eqnarray}
\alpha^* \alpha &=& \sum_k \bar{\alpha}_{ki} \alpha_{kj} |i \rangle \langle j |, \\
\omega_\lambda (\alpha^* \alpha) &=& \lambda \sum_k |\alpha_{k1}|^2 + 
(1-\lambda) \sum_k |\alpha_{k2}|^2 \geq 0.
\label{stateonalgebra}
\end{eqnarray}
So $\omega_\lambda$ are indeed states on $M_2(\mathbb{C})$.

{\it Cases $\lambda =0,1$}:

These are the easy cases.

If $\lambda=0$, then ${\cal N}_{\omega_\lambda} = {\cal N}_{\omega_0} \simeq \mathbb{C}^2$ and is 
spanned by elements of the form
\begin{equation}
\alpha = \begin{pmatrix}
	    	  \alpha_{11}  & 0   \\
   		  \alpha_{21} &  0
\end{pmatrix}.
\label{easycase}
\end{equation}
So $\widetilde{\cal N}_{\omega_0}$ is spanned by linear combinations of $|e_{11} \rangle$ and $|e_{21} 
\rangle$. Accordingly, the GNS Hilbert space ${\cal H}_{\omega_0} = 
\tilde{\cal A}/\widetilde{\cal N}_{\omega_0} \simeq \mathbb{C}^2$ has basis 
$|[e_{12}] \rangle, |[e_{22}] \rangle$. The representation $\pi_{\omega_0}$ is irreducible:
\begin{equation}
\pi_{\omega_0} (e_{ij}) |[e_{k2}] \rangle = \delta_{jk} |[e_{i2}] \rangle.
\end{equation}
Hence $\omega_0$ is pure.

We can find rank 1 density matrix $\rho_{\omega_0}$ associated with $\omega_0$ by finding the image of 
$|{\bf 1} \rangle$ in ${\cal H}_{\omega_0}$. We have
\begin{equation}
|{\bf 1} \rangle \rightarrow |[e_{22}] \rangle
\end{equation}
So
\begin{equation}
\rho_{\omega_0} = |[e_{22}] \rangle \langle [e_{22}]|
\end{equation}
Note that for $\lambda=0$, 
\begin{equation}
\langle [e_{22}]|[e_{22}] \rangle = \omega(e_{22}^* e_{22}) =1
\end{equation}
so that 
\begin{equation}
{\rm Tr}\, \rho_{\omega_0} =1.
\end{equation}

A similar situation prevails for $\lambda=1$. In this case,
\begin{eqnarray}
\widetilde{\cal N}_{\omega_1} &\simeq& \mathbb{C}^2 = {\rm Span}\, \{|[e_{12}] \rangle, |[e_{22}] \rangle \}, \\
\tilde{\cal A}/\widetilde{\cal N}_{\omega_1} &\simeq& \mathbb{C}^2 = {\rm Span}\, \{|[e_{11}] \rangle, 
|[e_{21}] \rangle \}
\end{eqnarray}
and
\begin{equation}
\rho_{\omega_1} = |[e_{11}] \rangle \langle [e_{11}]|, \quad {\rm Tr}\, \rho_{\omega_1} =1.
\end{equation}
As $\omega_1$ is pure, its entropy is zero:
\begin{equation}
-{\rm Tr} \rho_{\omega_1} \log \rho_{\omega_1} =0.
\end{equation}

{\it Case $0 < \lambda <1$}:

Here, from (\ref{stateonalgebra}), we can read off that
\begin{equation}
{\cal N}_\omega =\{ 0 \}.
\end{equation}
Hence
\begin{equation}
{\cal H}_{\omega_\lambda} = \{|m \rangle, m \in M_2(\mathbb{C}) \}
\end{equation}
The action of $M_2(\mathbb{C})$ on ${\cal H}_{\omega_\lambda}$ is
\begin{equation}
\pi_{\omega_\lambda} (\alpha) |m \rangle = |\alpha m \rangle
\end{equation}
Hence $\alpha$ acts separately on each column of $m$ and $\pi_{\omega_\lambda}$ splits into two 
two-dimensional irreducible representations:
\begin{equation}
\pi_{\omega_\lambda} = \pi^{(1)}_{\omega_\lambda} \oplus \pi^{(2)}_{\omega_\lambda}
\label{directsum}
\end{equation}
The corresponding decomposition of ${\cal H}_{\omega_\lambda}$ is
\begin{eqnarray}
{\cal H}_{\omega_\lambda} & = & \mathbb{C}^{(1)}_2 \oplus \mathbb{C}^{(2)}_2 \label{decomp}\\
\mathbb{C}^{(1)}_2 & = & \bigg\{ |\begin{pmatrix}
   						 m_{11}  & 0   \\
    						 m_{21} &  0
					     \end{pmatrix} \rangle \bigg\}  \label{uir1}\\
\mathbb{C}^{(2)}_2 & = & \bigg\{ |\begin{pmatrix}
   						 0  & m_{12}   \\
    						 0 &  m_{22}
					     \end{pmatrix} \rangle \bigg\} \label{uir2}
\end{eqnarray}
The direct sum in (\ref{directsum}) is the orthogonal direct sum.

An orthonormal basis for $\mathbb{C}^{(\alpha)}_2$ is 
\begin{eqnarray}
\mathbb{C}^{(1)}_2  & = & \Big\{ |\hat{e}_{11} \rangle = \frac{1}{\sqrt{\lambda}} | e_{11} \rangle, \quad  |\hat{e}_{22} \rangle = \frac{1}{\sqrt{\lambda}} | e_{21} \rangle  \Big\}\\
\mathbb{C}^{(2)}_2  & = & \Big\{ |\hat{e}_{11} \rangle = \frac{1}{\sqrt{1-\lambda}} | e_{12} \rangle, \quad  |\hat{e}_{22} \rangle = \frac{1}{\sqrt{1-\lambda}} | e_{22} \rangle  \Big\}
\end{eqnarray}

The irreducible representations $\pi^{(\alpha)}_{\omega_\lambda}$ are equivalent. Hence {\it the 
decomposition of ${\cal H}_{\omega_\lambda}$ into orthogonal direct sum is not unique.} For each 
$u \in SU(2)$, we can decompose ${\cal H}_{\omega_\lambda}$ into an orthogonal direct sum of invariant 
subspaces
\begin{equation}
{\cal H}_{\omega_\lambda} = \mathbb{C}^{(1)}_2 (u) \oplus \mathbb{C}^{(1)}_2 (u)
\end{equation}
with $\mathbb{C}^{(\alpha)}_2 ({\bf 1})$ being $\mathbb{C}^{(\alpha)}_2$ of (\ref{decomp}).
The orthonormal basis for $\mathbb{C}^{(\alpha)}_2 (u)$ are
\begin{eqnarray}
\mathbb{C}^{(1)}_2 (u)& : & \Big\{ |\hat{e}_{1r} \rangle u_{r1}, |\hat{e}_{2r} \rangle u_{r1} \Big\} \label{c1}\\
\mathbb{C}^{(2)}_2 (u)& : & \Big\{ |\hat{e}_{1r} \rangle u_{r2}, |\hat{e}_{2r} \rangle u_{r2} \Big\} \label{c2}
\end{eqnarray}
As emphasized by Sorkin, this ambiguity has serious consequences for the definition of entropy.

\section{Ambiguity in Entropy}

We divide this section into three parts: a) A general explanation on how entropy becomes non-unique, 
b) Illustrations from the $M_2(\mathbb{C})$ example from the previous section, and from an example using the 
harmonic oscillator, c) Circumstances when this 
ambiguity does not appear.

{\it a) General Discussion}

Let ${\cal A}$ be a unital $C^*$-algebra of observables, $\omega$ a state on ${\cal A}$ and 
${\cal H}_\omega$ the Hilbert space carrying the GNS representation $\pi_\omega$. Let us assume that 
$\pi_\omega$ is reducible and is the direct sum of several (say $N$) equivalent irreducible representations 
(IRR's):
\begin{equation}
\pi_\omega = \bigoplus_{r=1}^N \pi^{(r)}_\omega \label{GNSdecomp}
\end{equation}
Accordingly, ${\cal H}_\omega$ decomposes into an orthogonal direct sum of invariant subspaces:
\begin{equation}
{\cal H}_\omega = \bigoplus_{r=1}^N {\cal H}^{(r)}_\omega
\label{decomp2}
\end{equation}
The discussion is easily adapted when $\pi_\omega$ contains other inequivalent IRR's. 

Let $P^{(r)}$ be the projection operator to ${\cal H}^{(r)}_\omega$:
\begin{equation}
P^{(r)} P^{(s)} = \delta_{rs} P^{(r)}.
\end{equation}
Then the decomposition (\ref{decomp2}) means that 
\begin{equation}
\omega(\alpha) = \sum_r \omega (P^{(r)} \alpha P^{(r)}), \quad \alpha \in {\cal A}.
\label{statedecomp}
\end{equation}
Thus the restriction of $\omega$ to $\pi^{(r)}_\omega$ is $\Omega_r$, given by
\begin{equation}
\Omega_r (\alpha) = \omega (P^{(r)} \alpha P^{(r)}).
\end{equation}
$\Omega_r$ may not be a state, since $\Omega_r ({\bf 1}) = \omega(P^{(r)})$ may not be 1. It is a 
``weight". A pure state is thus 
\begin{equation}
\omega_r = \frac{1}{\Omega_r({\bf 1})} \Omega_r = \frac{1}{\omega({P^{(r)}})} \Omega_r
\end{equation}
and a decomposition of $\omega$ into pure states is
\begin{equation}
\omega = \sum \Omega_r({\bf 1}) \omega_r.
\label{omegadecomp}
\end{equation}
If $\big\{|e^{(r)}_i \rangle \big\}$ is an orthogonal basis for ${\cal H}^{(r)}_\omega$, then of course
\begin{equation}
P^{(r)} = \sum_i | e^{(r)}_i \rangle \langle e^{(r)}_i |
\end{equation}

But the decomposition (\ref{decomp2}) of ${\cal H}_\omega$ into invariant subspaces is not unique because 
of the degeneracy $N$ of the IRR's. Let $U(N)$ denote the unitary group of 
$N \times N$ matrices $u$ and let ${\bf 1}$ be its $N \times N$ unity. Then the orthonormal vectors 
\begin{equation}
|e^{(r)}_i (u) \rangle = u_{rs} |e^{(s)}_i \rangle
\end{equation}
for each fixed $u$ and $r$ span another invariant subspace under $\pi_\omega$ and 
\begin{equation}
{\cal H}_\omega = \bigoplus_r {\cal H}^{(r)}_\omega (u), \quad {\cal H}^{(r)}_\omega ({\bf 1}) = 
{\cal H}^{(r)}_\omega.
\end{equation}
Now ${\cal H}^{(r)}_\omega (e^{i \theta} {\bf 1})$ for any angle $\theta$ is also the same orthogonal direct sum.
So we have only $U(N)/U(1)$ worth of such distinct decompositions like (\ref{decomp2}) 
of ${\cal H}_\omega$.

Let $P^{(r)}(u)$ be the projectors to ${\cal H}^{(r)}_\omega (u)$.
\begin{equation}
P^{(r)}(u) = \sum_j |e^{(r)}_j (u) \rangle \langle e^{(r)}_j (u)|.
\end{equation}
Correspondingly we can write 
\begin{equation}
\omega(\alpha) = \sum \Omega_r (u|\alpha), \quad \Omega_r (u|\alpha) \equiv 
\omega(P^{(r)}(u) \alpha P^{(r)}(u))
\end{equation}
with
\begin{equation}
\Omega_r({\bf 1}| \cdot) = \Omega_r (\cdot)
\end{equation}
Now
\begin{eqnarray}
\Omega_r (u|{\bf 1}) & = & \omega (P^{(r)}(u)) \\
& = & \sum_s \Omega_s (P^{(r)}(u)) = \sum_s \omega (P^{(s)}({\bf 1}) P^{(r)}(u) P^{(s)}({\bf 1}))
\end{eqnarray}
Let us calculate $P^{(s)}({\bf 1}) P^{(r)}(u) P^{(s)}({\bf 1})$:
\begin{equation}
P^{(s)}({\bf 1}) P^{(r)}(u) P^{(s)}({\bf 1}) = \sum u_{rt} \bar{u}_{rt'} |e^{(s)}_i \rangle \langle e^{(s)}_i | \cdot 
|e^{(t)}_j \rangle \langle e^{(t')}_j | \cdot |e^{(s)}_k \rangle \langle e^{(s)}_k |
\end{equation}
with summation over $t,t',i,j,k$. This is
\begin{displaymath}
|u_{rs}|^2 P^{(s)}({\bf 1}).
\end{displaymath}
Thus we have the striking result
\begin{equation}
\Omega_r (u| {\bf 1}) = \sum_s |u_{rs}|^2 \Omega_s ({\bf 1})
\label{result}
\end{equation}

The decomposition (\ref{omegadecomp}) using pure states is now
\begin{equation}
\omega(\cdot) = \sum_r \Omega_r (u|{\bf 1}) \omega_r (u|\cdot), \quad \omega_r (u|\cdot) = 
\frac{1}{\Omega_r (u|{\bf 1})} \Omega_r (u|\cdot)
\end{equation}
Calling 
\begin{equation}
\lambda_r (u) = \Omega_r (u|{\bf 1})
\end{equation}
we find
\begin{equation}
\lambda_r (u) = \sum_s T(u)_{rs} \lambda_s ({\bf 1}), \quad {\rm where} \quad T(u)_{rs} = |u_{rs}|^2
\end{equation}

The entropy
\begin{equation}
S_\omega (u) = -\sum_r \lambda_r (u) \log \lambda_r (u) \label{uentropy}
\end{equation}
depends on $u$. We will have more to say about this in the next section.

{\it b) Examples}

i) $M_2({\mathbb C})$ for $0<\lambda<1$.

For $0<\lambda<1$, (\ref{decomp}) shows that ${\cal H}_{\omega_\lambda}$ is the direct sum of the two 
equivalent IRR's. Hence its decomposition into irreducible subspaces is not unique. This non-uniqueness can be 
parametrized by a unitary matrix $u \in U(2)$ as shown in (\ref{c1},\ref{c2}). For each such $u$, we generally 
get a different entropy as in (\ref{uentropy}).

In \cite{Balachandran:2012rw}, the entropy
\begin{equation}
S_{\omega_\lambda} ({\bf 1}) = - \lambda ({\bf 1}) \log \lambda({\bf 1}) - 
(1-\lambda({\bf 1})) \log (1-\lambda({\bf 1})) \label{oldentropy}
\end{equation}
was quoted as the entropy for this example. We can now understand that this result is not unique.

Still there is a special feature about (\ref{oldentropy}). If $S_{\omega_\lambda}(u)$ is minimized on $u$, we get 
(\ref{oldentropy}):
\begin{equation}
{\rm Min}_u S_{\omega_\lambda} (u) = S_{\omega_\lambda}({\bf 1}) \label{minentropy}
\end{equation}
We now show this result.

Since
\begin{equation}
\sum_r \lambda_r (u) = 1, \label{lambdasum}
\end{equation}
we have 
\begin{equation}
\delta \lambda_1 (u) + \delta \lambda_2 (u) = 0. \label{deltalambda}
\end{equation}
Hence 
\begin{equation}
\delta S_{\omega_\lambda} (u) = - \delta \lambda_1 (u) \log \frac{\lambda_1 (u)}{\lambda_2 (u)} \label{deltaS}
\end{equation}
This variation vanishes if
\begin{equation}
\lambda_1 (u) = \lambda_2 (u) \quad {\rm or} \quad \lambda_1 (u) = \lambda_2 (u) = 1/2.
\end{equation}
This corresponds to maximum entropy, as is well-known. It is not attainable if $\lambda_1 ({\bf 1}), 
\lambda_2 ({\bf 1}) \neq 1/2$ since if $\lambda_1 (u) = \lambda_1 (u)$, then $\lambda_i (u)$ are independent 
of $u$.

But the variation (\ref{deltalambda}) also vanishes if $\delta \lambda_1 (u)=0$. Thus using
\begin{equation}
|u_{11}|^2 + |u_{12}|^2 = 1, 
\end{equation}
we have 
\begin{equation}
\lambda_1 (u) = \lambda_1 ({\bf 1}) + |u_{12}|^2 [2 \lambda_1 ({\bf 1}) -1].
\end{equation}
So (\ref{deltaS}) also vanishes if 
\begin{equation}
\delta |u_{12}|^2 = |u_{12}| \delta|u_{12}| = 0 \label{deltau}
\end{equation}

That is (\ref{deltau}) vanishes if 
\begin{equation}
|u_{12}|=0
\end{equation}
which means that $|u_{11}|= |u_{22}|=1$.

For this choice, we get the value given in \cite{Balachandran:2012rw}. It is the minimum entropy.

ii) Simple harmonic oscillator

Let us consider a one-dimensional simple harmonic oscillator with annihilation and creation operators $a$
and $a^*$, vacuum $|0\rangle$ and the $n$-particle normalized state vector 
\begin{equation}
|n \rangle = \frac{{a^*}^n}{\sqrt{n!}} |0\rangle .
\end{equation}
The algebra of observables is the polynomial algebra in $a$ and $a^*$, subject to the relation $[a,a^*]=1$.

We want to show that the mixed state
\begin{equation}
 \omega = \lambda |0\rangle \langle 0| + (1-\lambda) |1\rangle \langle 1|, \quad 0<\lambda <1 \label{SHOstate} 
\end{equation}
gives two copies of the irreducible representation (IRR) of $a, a^*$. (By standard theorems, there is only 
one such IRR upto equivalence). We also want to show that the associated entropy depends on a unitary matrix 
$u$ unless $\lambda = 1/2$.

Let us first find the null space ${\cal N}_\omega$. If $\alpha \in {\cal N}_\omega$,
\begin{equation}
 \omega(\alpha^* \alpha) = 0
\end{equation}
or
\begin{equation}
 \lambda \langle 0| \alpha^* \alpha |0 \rangle + (1-\lambda) \langle 1|\alpha^* \alpha |1 \rangle = 0.
\end{equation}
Since both terms are positive, we get
\begin{equation}
 \alpha |0 \rangle = \alpha |1 \rangle =0.
\end{equation}

If 
\begin{equation}
 P = |0\rangle \langle 0| + |1 \rangle \langle 1| \equiv P^{(0)} + P^{(1)}
\end{equation}
is the projector to the subspace spanned by $|0\rangle$ and $|1\rangle$, this means that
\begin{equation}
 \alpha = \alpha ({\bf 1} - P).
\end{equation}
Conversely any $\beta ({\bf 1}-P)$, $\beta \in {\cal A}$ annihilates $|0\rangle$ and $|1\rangle$. Thus
\begin{equation}
 {\cal N}_\omega = {\cal A}({\bf 1}-P).
\end{equation}

Since
\begin{equation}
 {\bf 1} = ({\bf 1}-P) + P
\end{equation}
the component of $|{\bf 1}\rangle$ in ${\cal H}_\omega$ is
\begin{equation}
 |P + {\cal N}_\omega \rangle := | [P] \rangle.
\end{equation}
We see that $|[P] \rangle$ is a normalized vector:
\begin{equation}
 \langle [P] | [P] \rangle = \omega (P^2) = \omega(P) = 1
\end{equation}
so that a density matrix representing $\omega$ is 
\begin{equation}
 |[P]\rangle \langle [P]|. 
\end{equation}
Consider the subspaces
\begin{equation}
 {\cal H}_\omega^{(r)} = {\cal A} |[|r \rangle \langle r|] \rangle, \quad r=0,1
\end{equation}
of 
\begin{equation}
 {\cal H}_\omega = {\cal A} |[P] \rangle.
\end{equation}
Then 
\begin{eqnarray}
 {\cal H}_\omega &=& {\cal H}_\omega^{(0)} \oplus {\cal H}_\omega^{(1)}  \label{SHOspaces} \\
 \omega(\alpha^{*(1)} \alpha^{(0)}) &=& 0 \quad {\rm if} \quad \alpha^{(r)} \in {\cal H}_\omega^{(r)}. 
\end{eqnarray}
So the decomposition (\ref{SHOspaces}) is an orthogonal direct sum of invariant subspaces under ${\cal A}$. 
The representation $\pi_\omega$ of ${\cal A}$ on ${\cal H}_\omega$ is the direct sum of two IRR's. They are 
equivalent by general theorems on representations of the Weyl algebra.

Now
\begin{equation}
 \omega(P^{(0)}) = \lambda, \quad \omega(P^{(1)}) = (1-\lambda)
\end{equation}
so that the vectors 
\begin{equation}
 |e^{(0)} \rangle = \frac{1}{\sqrt{\lambda}} | [P^{(0)}] \rangle, \quad |e^{(1)} \rangle = 
\frac{1}{\sqrt{1-\lambda}} |[P^{(1)}] \rangle
\end{equation}
are normalized and orthogonal. The density matrix (\ref{SHOstate}) hence has the extremal decomposition
\begin{equation}
 | [P^{(0)}] \rangle \langle | [P^{(0)}]| + | [P^{(1)}] \rangle \langle | [P^{(1)}]| 
= \lambda \omega_0 + (1-\lambda) \omega_1, \quad \omega_r = | e^{(r)} \rangle \langle e^{(r)} |.
\end{equation}
with entropy
\begin{equation}
 S_\omega = -\lambda \log \lambda - (1-\lambda) \log (1-\lambda) \label{SHOentropy}
\end{equation}

As in the case of $M_2 (\mathbb{C})$, we can  decompose the space ${\cal H}_\omega$ into 
two new orthogonal invariant subspaces ${\cal H}_\omega^{(r)} (u)$ (for $u \neq ({\rm phase})\times{\bf 1}_{2}$), 
where
\begin{equation}
 {\cal H}_\omega^{(r)} (u) = {\cal A} |e^{(s)} \rangle u_{sr} := {\cal A} |e^{(r)} (u) \rangle, \quad u \in U(2).
\end{equation}
Then as in that case, if $\lambda \neq 1/2$, we get entropy $S_\omega (u)$ depending on $u$, 
(\ref{SHOentropy}) being the value at $u = {\bf 1}$.

{\it c) When do the ambiguities not appear?}

In the GNS construction, if an IRR of ${\cal A}$ appears only once, then there is a unique decomposition of a 
state $\omega$ into extremals and a unique entropy.

The second possibility is more subtle. It happens in non-abelian gauge theories with twisted bundles. 
We shall discuss it elsewhere where we discuss non-abelian monopoles, and molecules like $C_2 H_4$ based on 
non-abelian representations of discrete groups. It also happens for KMS states \cite{haagbook}.

The way it happens is as follows.

Consider (\ref{GNSdecomp}). For $N>1$, $\pi_\omega$ is reducible. The decomposition on the RHS is performed in 
all our examples using projectors $P^{(r)}$ to the subspaces ${\cal H}_\omega^{(r)}$. But it can happen that 
the algebra of observables ${\cal A}$ does not contain the projectors $P^{(r)}$. In that case, {\it even though 
$\pi_\omega$is the direct sum of equivalent IRR's}, the state has a unique decomposition.

Of course if we enlarge ${\cal A}$ by including including $P^{(r)}$, then $\omega$ has many extremal 
decompositions.

\section{On Stochastic Matrices}
The Markovian semi-group in classical probability theory arises as follows \cite{chruscinski}.

Let $\lambda = (\lambda_1, \cdots, \lambda_N)$ be a probability vector, with 
$\lambda_i \geq 0, \sum\lambda_i =1$. If $\lambda'$ is another probability vector, and the map 
$\lambda\rightarrow \lambda'$ is Markovian, the most general such map is by a stochastic matrix $T$:
\begin{equation}
\lambda'_r = T_{rs} \lambda_s, \quad T_{rs} \geq 0, \quad \sum_r T_{rs} =1. 
\label{markov}
\end{equation}
If $\det T \neq 0$, then $T^{-1}$ exists. But $T^{-1}$ is not a stochastic matrix, as seen from the example
\begin{equation}
T = \begin{pmatrix}
   1/2   &  1/2  \\
     1/4 &   3/4
\end{pmatrix}, \quad
T^{-1} = \begin{pmatrix}
     3 & -2   \\
     -1 & 2 
\end{pmatrix}.
\end{equation}
So the process (\ref{markov}) is irreversible.

In our case
\begin{equation}
T_{rs} = |u_{rs}|^2 \geq 0, \quad \sum_r T_{rs}=1
\end{equation}
so that $T$ is a stochastic matrix. It is actually doubly stochastic:
\begin{equation}
\sum_s T_{rs} =1.
\end{equation}

We can consider a Markovian evolution of $\lambda$ in ``time'', with 
\begin{equation}
\lambda_r (t) = T_{rs}(t) \lambda_s (0), \quad \lambda_s (0) = \lambda_s
\end{equation}
so that 
\begin{equation}
T(0) = {\bf 1}.
\end{equation}
There is an infinitesimal generator $L$ for $T(t)$. We have
\begin{eqnarray}
\frac{d \lambda_i (t)}{dt} &=& \sum_j \Big( \pi_{ij} \lambda_j (t) - \pi_{ji} \lambda_i (t) \Big) \\
&=& L_{ij} \lambda_j (t), \label{Keqn} \\
L_{ij} &=& \pi_{ij} - \delta_{ij} \sum_k \pi_{kj}
\end{eqnarray}
Note that 
\begin{equation}
\sum_i L_{ij} =0.
\end{equation}

Kolmogorov found the conditions on $L$ for the legitimacy of the equations of motion (\ref{Keqn}) 
on probabilities. They are
\begin{equation}
L_{ij} \geq 0, \quad {\rm if}\quad i \neq j, \quad \sum_i L_{ij} =0.
\end{equation}

In our case, 
\begin{equation}
T(u) T(v) \neq T(u+v)
\end{equation}
so that the map $u \rightarrow T(u)$ by no means preserves the group properties of $U(N)/U(1)$.

If $L$ is independent of $t$, then 
\begin{eqnarray}
T(t) &=& e^{-t L} , \\
T(t+s) &=& T(t) T(s), \quad {\rm for} \quad t,s >0
\end{eqnarray}
generates a semi-group: it is only a semi-group in general since $e^{+t L}$ may not be a stochastic matrix.

In our case where $T_{rs} = |u_{rs}|^2$, even if $t \rightarrow u(t) \equiv e^{-i t \ell}$ is a one-parameter subgroup,
$\frac{dT(t)}{dt}$ does not come out as an expression in $T_{ss'}(t)$ and $\frac{dT_{ss'}(t)}{dt}$:
\begin{equation}
\frac{d}{dt} T_{rs}(t) = -i \bar{u}_{rs} (u \ell)_{rs} + i (\bar{\ell} \bar{u})_{rs} u_{rs}.
\end{equation}

It is easy to identify $T$ with a completely positive map on density matrices, that is, write $\rho(u)$ in the Kraus 
form as 
\begin{equation}
\rho (u)  =  \sum_r \Lambda^{(r)} (u) \rho({\bf 1}) \Lambda^{(r)}(u)^{\dagger}.
\end{equation}

Thus
\begin{equation}
 {\bf 1} = \sum_r P^{(r)} (u)
\end{equation}
implies that
\begin{equation}
 \rho (u) = \sum_r |P^{(r)}(u) \rangle \langle P^{(r)}(u)|.
\end{equation}
Now
\begin{equation}
 \langle P^{(r)}(u)|P^{(s)}(u) \rangle = \delta_{rs} \omega (P^{(r)}(u)) = \delta_{rs} \lambda_r (u).
\end{equation}
Hence
\begin{equation}
 \Lambda^{(r)}(u) = \frac{1}{\lambda_r ({\bf 1})} |P^{(r)}(u) \rangle \langle P^{(r)}({\bf 1})|.
\end{equation}

Note that 
\begin{equation}
 \sum_r {\rm Tr} \Lambda^{(r)}(u) \rho({\bf 1}) \Lambda^{(r)}(u)^\dagger =1
\end{equation}
even though 
\begin{equation}
 \sum_r \Lambda^{(r)}(u)^\dagger \Lambda^{(r)} (u) = 
\sum_r \frac{\lambda_r(u)}{\lambda_r({\bf 1})^2} |P^{(r)}({\bf 1}) \rangle \langle P^{(r)}({\bf 1})| \neq {\bf 1}.
\end{equation}

The inverse to the map $\rho({\bf 1}) \rightarrow \rho(u)$ will not exist if a $\lambda_r(u)$ vanishes.

%

\section{Quantum Entropy and its Increase}

The following theorem is due to Kubo \cite{Sorkin:1997ja}. It is the simple statement 
that the entropy
\begin{eqnarray}
\hat{S}(\lambda) & = & \sum_i \hat{s}(\lambda_i) \\
\hat{s}(x) & = & - x \log x/x_0, \quad \hat{s}(x_0) = 0,
\end{eqnarray}
increases under the stochastic map:
\begin{equation}
\hat{S} (T \lambda) \geq \hat{S}(\lambda).
\end{equation}
This follows from the fact that $\hat{s}(x)$ is concave for $x \in [0,1]$:
\begin{equation}
\hat{s}''(x) <0 \quad {\rm for} \quad x \in [0,1].
\end{equation}

What is its relevance for our work? It is that the map $\lambda \rightarrow T(u)\lambda$ is stochastic as 
we proved. It means that if we change the representation of $\omega$ in extremals using a $u$, the entropy 
increases.

That means for example that if we put a dynamics on $u$, possibly unitary, it will project to an 
irreversible dynamics on $\lambda$ with non-decreasing entropy under evolution. 

One can see from the formulae that the maximum entropy is reached in the most disordered state where 
$\lambda_i = \lambda_j = 1/N$. If $N = \infty$, it may never be reached.

We will discuss realistic models for entropy increase in papers under preparation.

\section{Acknowledgments} 
It is a pleasure to thank Rafael Sorkin for discussions on entropic 
matters, and in particular for pointing out its possible ambiguities. We would also like to thank Manolo Asorey, 
Darek Chru\'sci\'nski, T. R. Govindarajan, Beppe Marmo and Andres Reyes-Lega for discussions.
APB is supported by the Institute of Mathematical Sciences, Chennai. ARQ is supported by CNPq under 
process number 307760/2009-0.

\end{document}